\documentclass[%
 reprint,
 amsmath,amssymb,
 aps,nofootinbib,
]{revtex4-2}
\usepackage{xcolor}

\usepackage{graphicx}
\usepackage{dcolumn}
\usepackage{bm}
\usepackage{hyperref}
\usepackage[normalem]{ulem}

\begin{document}


\title{Proton Synchrotron, an explanation for possible extended VHE gamma-ray activity of TXS 0506+056 in 2017}

\author{Sunanda}
 \email{sunanda@iitj.ac.in}
\author{Reetanjali Moharana}%
 \email{reetanjali@iitj.ac.in}
\affiliation{Department of Physics, Indian Institute of Technology Jodhpur, Karwar 342037, India.
}%
\author{Pratik Majumdar}%
\affiliation{Saha Institute of Nuclear Physics, HBNI, Kolkata, West Bengal 700064, India
}


\date{\today}

\begin{abstract}
TXS 0506+056, a source of the extreme energy neutrino event, IceCube-170922A, was observed on 22 September 2017. The Fermi-LAT detector reported high energy (HE) $\gamma$-ray flare between 100 MeV and 100 GeV starting from 15 September 2017 from this source. Several attempts to trace the very high energy (VHE) gamma-ray counterparts around the IceCube-170922A resulted in no success. Only after 28 September, the Major Atmospheric Gamma-ray Imaging Cherenkov (MAGIC) telescopes observed the first VHE gamma-rays from the blazar above 100 GeV. The $\sim$ 41 hr survey resulted in VHE-$\gamma$ ray activity till 31 October 2017.  Here we propose the extended GeV $\gamma-$rays can be explained by taking two production channels, electron synchrotron self Compton and proton synchrotron for HE and VHE emissions, respectively. The 45 days of VHE emission from the peak of the HE-flare can be explained with $ {L_p'}{\simeq}10^{47}$ erg/sec in the jet frame and magnetic field of 2.4 G, consistent with the $\textup{L}_{Edd}$ for a blackhole mass $5\times 10^{9} \textup{M}_\odot$ . 

\end{abstract}

\maketitle


\section{\label{sec:intro}Introduction}
Observational correlation of an  $>290$ TeV, extremely high energy (EHE) neutrino event from the flaring direction of TXS 0506+056 blazar on 22 September 2017 by IceCube (IceCube-170922A) offers a potential breakthrough in understanding blazars and their hadronic emissions. The follow-up multi-wavelength study for the neutrino event has enabled a deep understanding of the hadronic models for blazars. Such a synchronized observation over the broad energy range also has the potential to reassess the current understanding of different blazar emission models. 

A Follow-up Study of IceCube-170922A by $Fermi$ Large Area Telescope (LAT) reported a gamma-ray (0.1-300 GeV) excess of $3.6 \pm 0.5 \times 10^{-7}$ photons cm$^{-2}$ sec$^{-1}$ (statistical error only) within time period 15 to 27 September 2017 from TXS 0506+056 blazar, establishing the temporal correlation of the neutrino event \cite{Fermi_atel}. AGILE also confirmed $\gamma$-ray activity above 100 MeV from the source\cite{AGILe} with the maximum emission occurring one or two days before the neutrino event. Subsequently, a multi-wavelength campaign started for TXS 0506+056. An excess in the X-ray spectra from $2.32^{0.33}_{-0.29}$ to $2.5^{0.23}_{-0.12}$ was observed on 27 September with 5ks observation by Swift-XRT. ASAS-SN reported a rise of $~0.5$ mag in V-band \cite{ASAS}. Table.\ref{tabtime} lists the details of the follow-up observations of the blazar by different detectors around the globe.

A sub-GeV and TeV gamma-ray search was also followed-up after the neutrino event. High Energy Stereoscopic System (H.E.S.S) telescope searched around the region of the neutrino event for two consecutive nights and found no significant very high energy (VHE) emission \cite{HESS}. A similar search was carried out by VERITAS and no VHE source was found within $3.5$ degree of the neutrino event \cite{veritas}. VERITAS also looked for VHE emission from the blazar TXS 0506+056 between 28 and 30 September, and a non-detection resulted in an upper limit of $6.8 \times 10^{-12}$ cm$^{-2}$ sec$^{-1}$ at 99\% CL above 160 GeV\cite{veritas}. The High-Altitude Cherenkov (HAWC) detector examined its observation from 15  to 19 September, then from 21 to 27 September and 23-24 September, and found no evidence of emission above 1 TeV \cite{Hawc}. Major Atmospheric Gamma Imaging Cherenkov (MAGIC) Telescopes reported the first VHE observation from the blazar under good weather conditions with $374 \pm
62$ excess photons, up to energy 400 GeV marking a $6.2\sigma$ excess over expected background between 28 September and 3 October \cite{Magic}. {\cite{2018ApJ...863L..10A} has reported VHE activity between 1.3 to 40 days after the neutrino event from the source by monitoring nearly 41 hr with MAGIC telescopes. During this time two periods of enhanced VHE emission were observed, one between MJD 58029-58030 (2017 October 3–4) and the second on MJD 58057 (2017 October 31). A similar activity is also observed during the time period of 2018 December 1 and 3 (MJD 58453 and 58455) while monitoring the time period between 2017 November and 2019 February, a total of nearly 79 hr \cite{2022ApJ...927..197A}. The light curve of HE observed by Fermi-LAT in \cite{2022ApJ...927..197A} also reports a flare in the time period 58087-58091 MJD.}

Dedicated attempts have been made by several groups to explain the multi-wavelength flare and the neutrino event from TXS 0506+056 with leptonic and photo-hadronic emission \cite{Gao:2018mnu,Keivani:2018rnh,Cerruti:2018tmc,Oikonomou:2019djc,Petropoulou:2019zqp}, and also proton-proton interactions \cite{Murase:2018iyl,Sahakyan:2018voh,Liu:2018utd,Banik:2019jlm,Banik:2019twt}. Time-dependent modelling of the TXS 0506+056 multi-wavelength flare has been done by \cite{Gao:2018mnu} with synchrotron and synchrotron self Compton emission (SSC), and the neutrino event with photo-hadronic interactions. This scenario requires super-Eddington jet power to explain the neutrino event. Additionally, the X-ray emission constrains the photo-hadronic model of neutrino production. \cite{Keivani:2018rnh} considered synchrotron and external Compton (EC) emission of relativistic electrons to explain the multi-wavelength spectrum of the flare and radiatively subdominant hadronic emission to explain the IceCube-170922A event.

The detailed follow-up observations suggest approximately 45 days of extended VHE activity following the neutrino-correlated Fermi-LAT HE flare. We explain this extended period of VHE activity with proton synchrotron. We explain the multi-wavelength observation of TXS 0506+056 blazar with a standard single-zone lepto-hadronic model.  Specifically the several hundred MeV HE emissions with electron SSC and the MAGIC VHE events with proton synchrotron. We also explain the neutrino event with proton-proton interaction inside the blob.


Section \ref{sec:model} discusses the multi-wavelength data collected for different detectors and the Fermi-LAT data analysis for TXS 0506+056. This section also details the lepto-hadronic modelling of the source. Section \ref{sec:res} is dedicated to the resulting time delay from the lepton and hadronic model. 

\section{\label{sec:model} Multi-wavelength Modeling of TXS 0506+056}
We collected the synchronized multi-wavelength data for TXS 0506+056 with the IceCube-170922 neutrino event; details are listed in table-\ref{tabtime}. We analyzed the Fermi-LAT data for the time of the multi-wavelength search of the source. We model the emissions till the HE gamma-rays with electron synchrotron and SSC, whereas the VHE gamma-ray events with proton synchrotron in the jet of TXS 0506+056.

\begin{table}[h!]
	\centering
	\caption{Multiwavelength observations of TXS 0506+056 blazar during different epochs}
	\label{tabtime}
\begin{tabular}{lccr} 
   \hline
	Detector & Epoch (MJD)& Reference \\
        \hline
        VLA &$ 58031,58032,58035,58038$ &\cite{VLA-2017}\\
        
        $Kanata$& $58019,58020,58021$ & {\cite{kanata-2017}}\\
        Swit UVOT& $ 58019$ & \cite{Swift-XRT}\\
        
        Nustar& $58025$ &\cite{Nustar} \\
        
        Swift XRT & $58023$, $58026$ & \cite{Swift-XRT}\\
       
   Fermi-LAT & $58011$ {to} $58023$ & \cite{Fermi_atel}\\
   
   \hline\hline
   VERITAS & $58018$, $58019$, $58024$, $58026$ &\cite{veritas} \\
   H.E.S.S & $58019$,$58020$ $58021$ & {\cite{HESS}} \\
   MAGIC & $58024$ {to} $58029$ & {\cite{Magic}} \cite{2018ApJ...863L..10A}\\
        
\hline
\end{tabular}
\end{table}
\subsection{HE gamma-ray light-curve of TXS 0506+056}
\begin{figure}[h!]
    \centering
    \includegraphics[width=9cm]{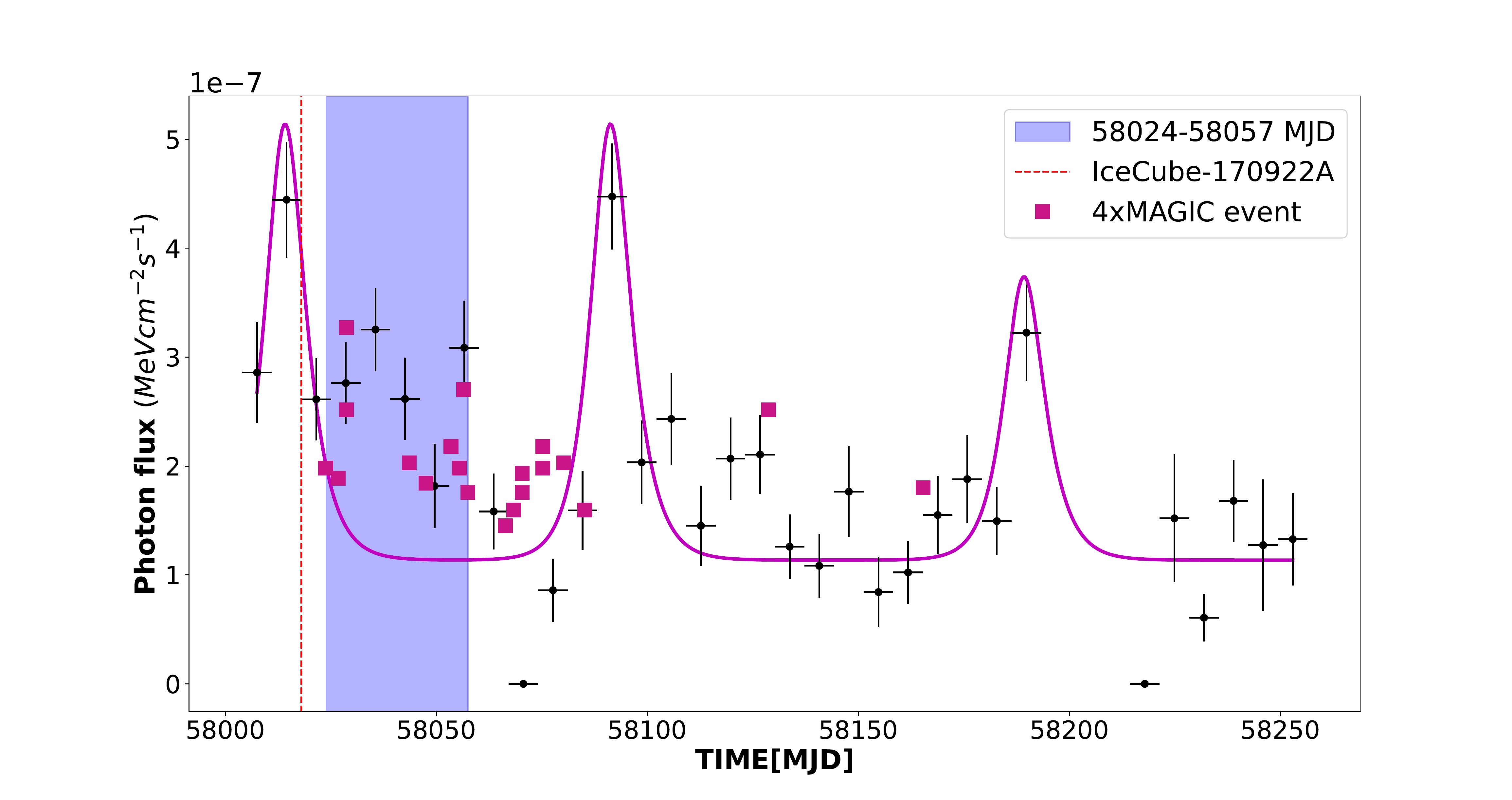}
    \caption{The Fermi-LAT light curve of TXS 0506+056 blazar for energy range 1 to 300 GeV between MJD 57997-58253 with 7-day bin. The (red) dotted line shows the correlated IceCube-170922A neutrino event within the HE flaring episode 58008-58021 MJD. The shadow region indicates the time period of VHE activity observed by MAGIC. The square points represent the MAGIC events reported in \cite{2022ApJ...927..197A}.}
    \label{fig:lc}
\end{figure}

Using the Fermi-LAT light curve repository (LCR)\footnote{https://fermi.gsfc.nasa.gov/ssc/data/access/lat/\\LightCurveRepository/index.html} we generated the light curve for TXS 0506+056 (4FGL J0509.4+0542) for 7 days time bin with a region of interest (ROI) $10^{\circ}$ centring the source, using Fermipy Tool \footnote{https://fermipy.readthedocs.io/en/0.14.1/\_modules/\\fermipy/lightcurve.html} for MJD 57997-58253. The light curve is shown in figure \ref{fig:lc} for the energy bin 100 MeV to 300 GeV. We observed three different activities within this time period. 

To study the time period of the flares we fitted each of the first two activities by a sum of exponential functional form\cite{2010ApJ...722..520A} which depicts the rising and decay time of the flare. The functional form of the fitted function is as follows 

\begin{equation}
\label{eq:pythagorean}
 F(t)=2F_0 \left(\frac{exp(t_0-t)}{T_r}+\frac{exp(t-t_0)}{T_d}\right)^{-1},   
\end{equation}
where $F_0$ is photon flux at time $t_0$, T$_r$ and T$_d$ represent the rise and decay times of the peak respectively.\\

\begin{figure}[h!]
    \centering
    \includegraphics[width=9cm]{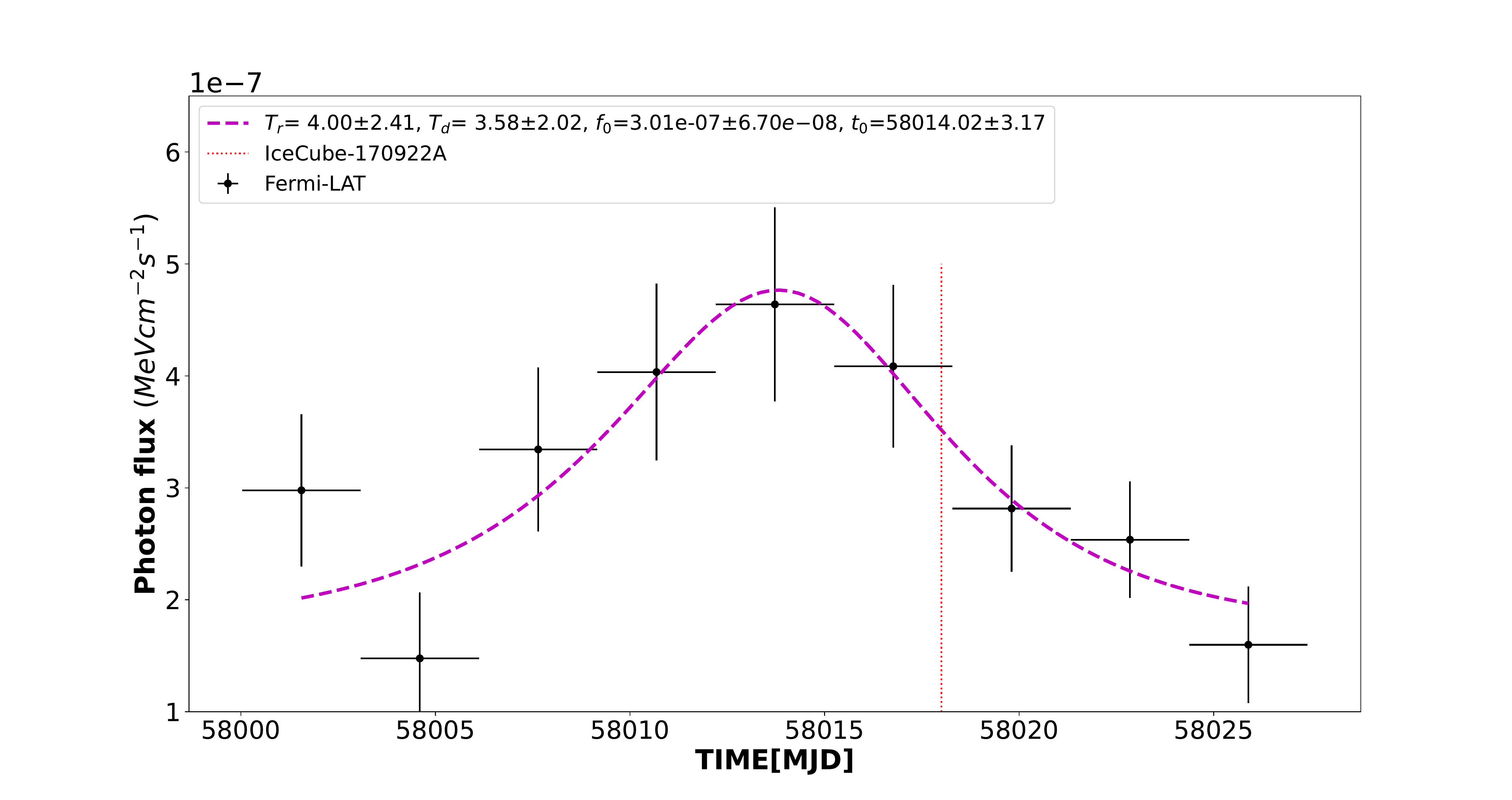}
    \caption{The Fermi light curve of TXS 0506+056 for the flaring episodes 58008-58021 MJD in the energy range .1-300 GeV with 3 days time bin. The correlated IceCube-170922A neutrino event is shown by the red line falls in flaring episodes. The rising and decay time of the flare has been calculated by the fitting function\ref{eq:pythagorean}.
}
    \label{fig:lc_first}
\end{figure}

We obtained the rising and decay time as $4.00\pm2.41$ and $3.58\pm2.02$ days respectively with peak time at, $t_0 = 58014.02\pm 3.17$ for the first flare during period 58008-58021 MJD. Hence the flare means the time period is 14 days, as it is $2(T_d+T_r)$). Figure \ref{fig:lc_first} shows the first flare with 3-day bin events as observed by Fermi-LAT. The fitted function for the flare is shown with dashed lines. This flare is correlated with the IC170922-A, the time of the neutrino event is shown with a dotted line in figure \ref{fig:lc_first}. Similarly, the rise and decay time obtained for the second flare in the time period 58087-58091 MJD (30 November 2017 to 4 December 2017) is $1.28\pm.25$ and $2.95\pm.25$ days respectively. 

Figure \ref{fig:lc} also shows the VHE observations by MAGIC with square points. The two activities reported in \cite{2018ApJ...863L..10A} are shown within the shadow region, within the time period 28 September to 31 October. Since the activity is followed by the first HE Fermi-LAT flare, we correlate these VHE activities with this HE flare. Though \cite{2022ApJ...927..197A} reported another VHE activity from 1 to 3 December 2018 observed After the first HE flare.


\subsection{Fermi-LAT data analysis for SED}

\begin{figure}[h!]
    \centering
    \includegraphics[width=9.2 cm]{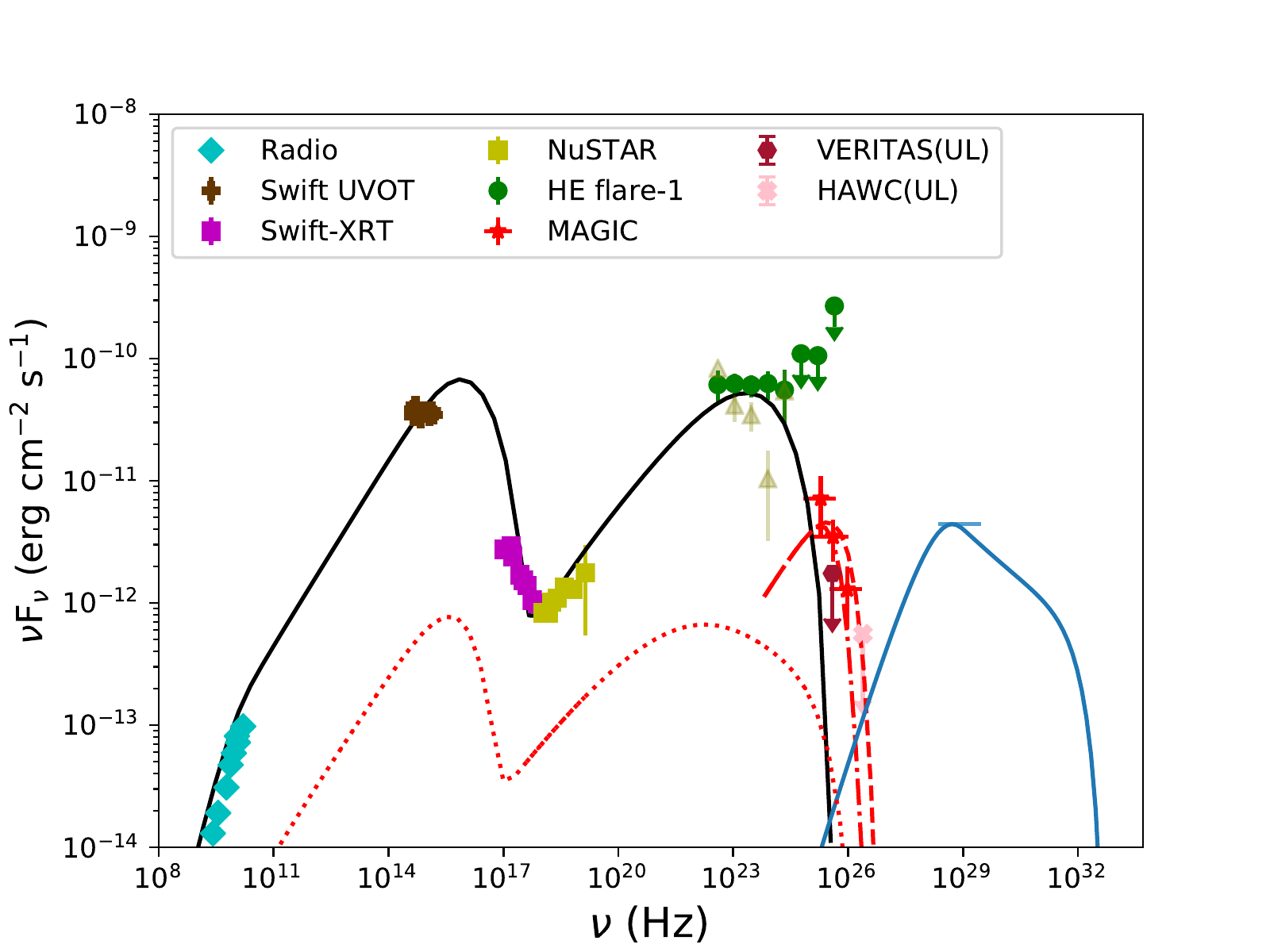}
    \caption{Modeling of the Multi-wavelength observations of TXS 0506+056 blazar as a follow-up of IceCube-$170922A$. The (cyan) rectangle points represent radio observations by VLA, (brown) plus points show the optical, (purple) squares for Swift-XRT and (yellow) squares for hard X-rays by Nustar. The (green) circles and (lemon) triangles represent HE-$\gamma$-rays  by Fermi-LAT, and the (red) star shows the VHE $\gamma$-rays by MAGIC. SED Modeling: the electron synchrotron and SSC are shown with a (black) solid line and the proton synchrotron with a (red) dashed line. The gamma-ray emission after EBL contribution is shown with the (red) dot-dashed line.}
 
    \label{fig:sed_mod}
\end{figure}

We have used the Fermi-LAT spectral energy distribution (SED) for the time period MJD 58008 to MJD 58021 or for the HE-flare. We extract the SED using Fermipy\footnote{https://fermipy.readthedocs.io/en/latest/} tool. First, we collected the data from the LAT data server for the source TXS 0506+056 blazar within 1$^{\circ}$ radius ROI centred at the location of the source for the SED analysis. The point spread function (PSF) will be unaffected for low-energy photons, as there is no point source within 1$^{\circ}$ region of the blazar listed in the fourth catalogue (4FGL). The unbinned maximum-likelihood analysis was performed with fermitool of version 2.2.0 and the instrument response function $\textup{P8R3{\_}SOURCE{\_}V3}$ $\textup{P8R2{\_}SOURCE{\_}V6}$. For our analysis, all the parameters of the Galactic diffuse model $\textup{({gll}{\_}{iem}{\_}v07.fits)}$ and isotropic component $\textup{{iso}{\_}{P8R3}{\_}{SOURCE}{\_}{V3}{\_}{v06}.{txt}}$ are kept free within 1$^{\circ}$ of ROI. In the event section Front+Back event type (evtype=3), evclass =128, and a zenith angle cut of 90$^{\circ}$ are applied based on 8 pass reprocessed source class. We have done the same analysis for the time period MJD 58020 to MJD 58030 (24 September to 4 October). The SED points extracted are shown in figure \ref{fig:sed_mod} with circle (green) and up triangle (grey) points for the two periods, respectively. 
\subsection{Lepton Modeling for HE gamma-rays}
We modelled the SED of the TXS 0506+056 source for the HE-flare period
with the emissions from a single zone blob of radius, R. The accelerated electrons follow a spectral form $E_e'^{-\alpha_e}$ ($'$ - represent jet frame) in the blob. The photon emissions in UV, optical, and soft X-ray energies from the source are explained with electron synchrotron in the blob magnetic field, B. The electron synchrotron cooling time is, 
\begin{equation}
    t^{obs}_{syn,e}\simeq{7.91\times10^{-3}}\;{\delta^{-1}_{16}}\; {B^{-2}_{2.4}}\;{E'^{-1}_{e,10}}\; \textup{days}
\end{equation}
where ${E'_{e,10}}= E'_{e}/10^{10} eV$. The hard X-rays and HE gamma-rays are modelled with time-dependent SSC by the relativistic electrons using GAMERA-package\cite{Hahn:2015hhw}. The SED modelling to explain the data points are presented in figure \ref{fig:sed_mod}. 
The modelling parameters details are listed in table \ref{tab:parmod}. 

\subsection{Modeling Proton Synchrotron for VHE gamma-rays}

Relativistic protons in the blob also follow the non-thermal flux $E_p'^{-\alpha_p}$ with exponential cut-off. We consider the acceleration of relativistic protons in the same region up to the maximum energy ${E_{p, max}^{\prime}}$.
{\begin{equation}
t^{obs}_{acc}\simeq{1.8}\;{E'^{p}_{19}} \;{\delta^{-1}_{16}}\;{B^{-1}_{2.4} \; {\eta_{4}}}\; \textup{days},  
\end{equation}
where gyro-factor $\eta_4 = \eta/4$.}

We modelled the VHE emissions observed by MAGIC with proton synchrotron in the blob with the magnetic field, B. The synchrotron cooling time for proton is, 
\begin{equation}
t^{obs}_{syn,p}\simeq{7.52}\; {B^{-2}_{2.4}}\;{\delta^{-1}_{16}}\;{E'^{-1}_{p,19}}\; \textup{days},
\label{tpsyn}
\end{equation}
where ${E'_{p,19}}=E'_p/10^{19} eV$. Details of the parameters used for modelling are listed in Table \ref{tab:parmod}.

\setlength{\tabcolsep}{1em} 
\begin{table}[h!]
	\centering
	\caption{The Parameters used in the lepto-hadronic model for the SEDs of TXS 0506+056 blazar}
	\label{tab:parmod}
	\begin{tabular}{lccr} 
	   \hline
		Parameters& Values \\
            \hline
            z & $.336$ \\
        d(pc) & $1.79\times10^9$\\
        \hline
            $\delta$ & $16$\\
            $\Gamma_{j}$ & $8$\\
            B(G) & $2.4$ \\
                    $R' $ (cm)& $1.23\times10^{16}$ \\
\hline
            $\alpha_{e}$ & $1.65$\\
            
            ${\gamma_{e,min}'}$ & ${4500}$ \\
            
            ${\gamma_{e,max}'}$& ${2.1\times10^4}$\\
             
   \hline     
         $\alpha_p$ &{2.01}\\
       $E_{p,min}'${(eV)}& $10^{14}$\\
       $E_{p}'^{b}${(eV)}& $2.95\times10^{18}$\\
       
        $E_{p,max}'${(eV)}& $4.6\times10^{19}$\\
    \hline
    \hline
             $L'_e({erg}\;{s^{-1}})$ & $2.9\times10^{43}$\\
        $L'_p({erg}\;{s^{-1}})$ & $5.9\times10^{47}$\\
    
   \hline
	\end{tabular}
\end{table}
\subsection{Neutrinos from proton-proton interaction}
We modelled the neutrino flux observed by IceCube during the HE-flare from TXS0506+056 with $pp$ interaction following \cite{Banik:2019jlm}. The gas density obtained by \cite{Banik:2019jlm} to model IC170922-A is $n_H=1.68\times 10^6 \, {cm^{-3}}$. However, the time required for $pp$ interaction in the observer frame is, 
\begin{equation}
t^{obs}_{pp}\simeq{3.52}\; {\delta^{-1}_{16}}\; {n_{H,10^{8}}^{-1}}\;  \left(\frac{\sigma_{pp}(E'_p)}{\sigma_{pp} (10^{16}\textup{eV})}\right)^{-1} \; \textup{days},
\label{tpp}
\end{equation}
where $n_{H,10^{8}} = n_H / (10^8 \;cm^{-3} )$, $\delta_{16}=\delta/16$, we have used the proton-proton cross-section $\sigma_{pp} (E'_p)$ from \cite{2006PhRvD..74c4018K}. And with the parameters obtained by \cite{Banik:2019jlm}, it takes more than 160 days (beyond the HE-flare) in the observer frame to achieve the IC170922-A flux. Here we propose that ambient density is not a constant value but a varying parameter, precisely the ambient decreases over time with index $\alpha_a$,
\begin{equation}
n_H= n_{H,0} \, t^{-{\alpha_a}} \; \textup{cm$^{-3}$},
\label{amb}
\end{equation}
One can expect such a variation due to the dynamics of the jet and blob. We emphasize that accelerated protons interact with the ambient; hence for fitting, we consider the evolution of the ambient over accelerated time.
Hence, the ambient density becomes,
\begin{equation}
n_H \simeq \left(1.8\times 10^{6} \right)^{-\alpha_d} n_{H,0}\,{\eta_{4}}^{-\alpha_d}{B^{\alpha_d}_{2.4}} {E'^{p}_{19}}^{-\alpha_d}\textup{cm$^{-3}$}.
\label{density}
\end{equation}
The parameter $\alpha_a$ and $n_{H,0}$ values are obtained from fitting the proposed model. 

\section{\label{sec:res}Results \& Discussion}
The light curve of TXS 0506+056 at a redshift $z=0.336$ observed by Fermi-LAT within the energy range 0.1 GeV to 300 GeV suggests a HE-flare from MJD 58008 to MJD 58021. IceCube-170922A neutrino event occurred at the same flaring episode. Follow-up observations by ground-based Imaging Atmospheric Cherenkov Telescopes (IACTs) like VERITAS, H.E.S.S reported non-detection with upper limits $1.2 \times 10^{-11}$ cm$^{-2}$ s$^{-1}$ and $7.5 \times 10^{-12}$ cm$^{-2}$ s$^{-1}$ respectively above $>175$ GeV.
A total of five hours of observations centring TXS 0506+056 between September 28 and 30 by VERITAS also reports no evidence of gamma-ray emission at the blazar location \cite{veritas}. Whereas MAGIC reported VHE activity starting its observation from 28 September to 31 October 2017. Hence the multi-wavelength model should incorporate the extended VHE emission for 45 days from the peak of the HE flare. We explain this extended time of VHE emission with proton synchrotron.


The lepton modelling for the HE-flare requires a magnetic field of B = $2.4$ Gauss in an R$' = 1.23\times 10^{16}$ cm blob with doppler $\delta= 16$ and Lorentz factor $\Gamma = 8$. The observed HE-flare time, 14 days is more than the light crossing time,  $\frac{R' (1+z\,)}{c \delta}$.The electron spectra follows a spectral index $\alpha_e = 1.65$ within energies $\gamma'_{e,min} = 4500$ to $\gamma'_{e,max} = 2 \times 10^4$. This results a jet frame electron luminosity, $L'_{e} = 2.9 \times 10^{43}$ erg/sec. The electron synchrotron and SSC model are shown in figure \ref{fig:sed_mod} with a solid (black) line. 

\subsection{\bf Proton Synchrotron as a probe for the extended VHE events:} 
We calculated the neutrino flux and the VHE energy photons with the proton-proton interaction and proton synchrotron in the same magnetic field of 2.4 G, respectively. The modelling resulted with a proton spectrum index, $\alpha_p=2.01$ within energy $E'_{p,min}= 10^{14}$ eV to $E'_{p,max}= 5\times 10^{19}$ eV. The $E'_{p,max}$ is a few times more than the Hillas criterion, $E_{p,max} = qRB$, q charge of an electron. However, as suggested by \cite{Ptitsyna:2008zs}, the maximal energy of the particles in the jet of a blazar is determined by radiation losses rather than by the Hillas condition following,
\begin{equation}
    E'_{p,max} = 3.7 \times 10^{19} \textup{eV} \left(\frac{M_{BH}}{10^8 \textup{M}_\odot}\right)^{3/8}.
    \label{epmax}
\end{equation}


\begin{figure}[h!]
    \centering
       \includegraphics[width=9cm]{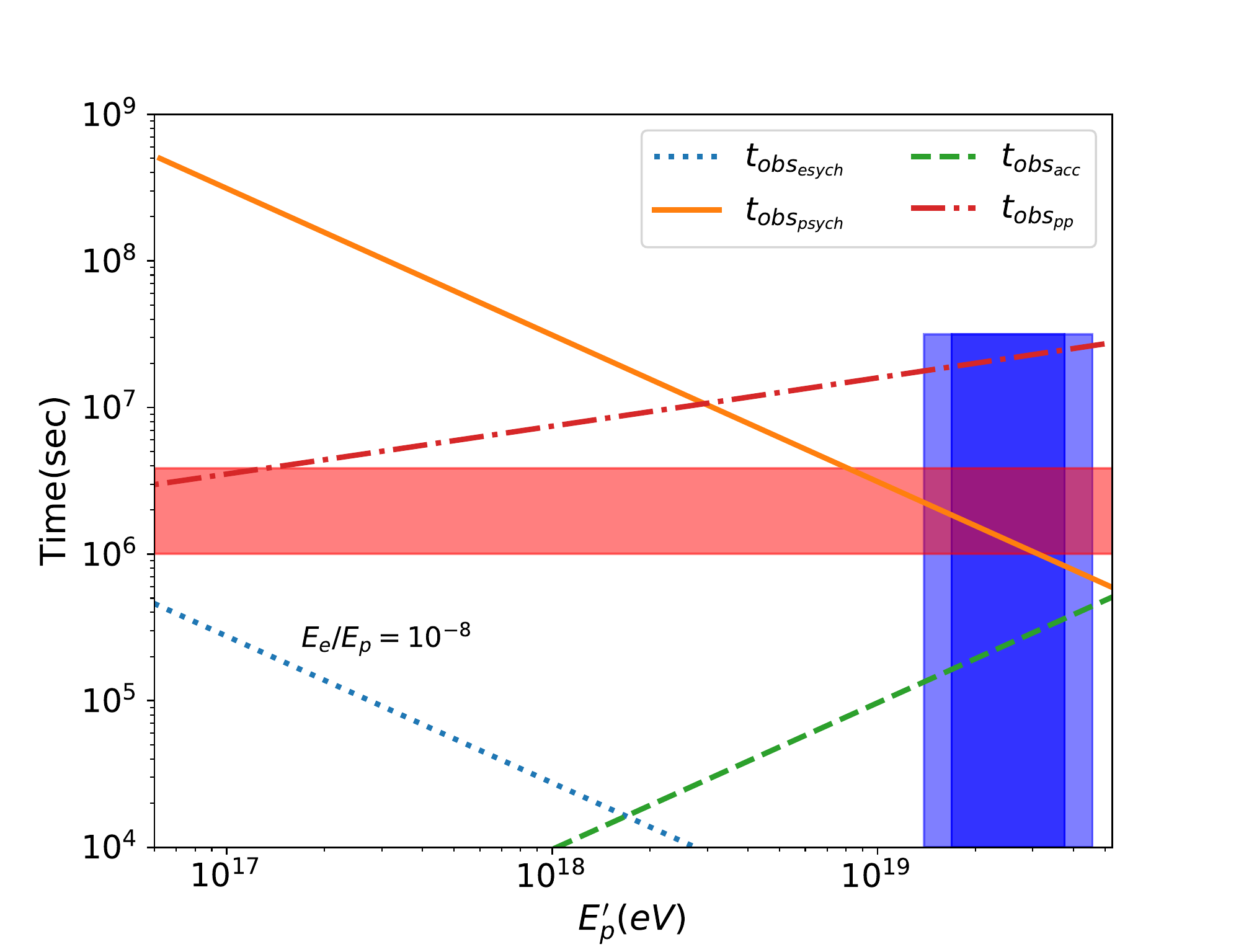}
    \caption{The orange dash line is the proton synchrotron cooling time and the blue dash line delineates the electron synchrotron cooling time for $E_e{=}E_p\times10^{-8}$ and  the blue shaded region marks the proton energy region for which proton synchrotron cooling time between 11 to 44 days in the observer frame.}
    \label{fig:mod_time}
\end{figure}
By considering the $M_{BH}$ of TXS 0506+056 as, $5 \times 10^{9}\textup{M}_\odot$ \cite{Gao:2018mnu}, our $E'_{p,max}$ is consistent with equation \ref{epmax}. The total jet frame proton luminosity is $L'_{p} = 5.9 \times 10^{47}$ ergs/sec, which is again consistent with the Eddington luminosity of black-hole mass taken for TXS 0506+056. The proton synchrotron gamma-ray emission is shown in figure \ref{fig:sed_mod} with dashed (red) lines. The VHE  $\gamma$-ray emissions from the source are attenuated due to interaction with Extragalactic background light (EBL). We calculated this suppression using \footnote{http://www.astro.unipd.it/background/} for the source at a distance $d= 1.79 \times 10^9$ pc.  The gamma-ray emission after this contribution is shown with dot-dashed (red) lines in figure \ref{fig:sed_mod}.  

The observer frame proton synchrotron (solid), and the proton-proton interaction(dash-dot) time scale over the energy range $E'_{p,min}$ to $E'_{p,max}$ in the blob magnetic field of 2.4 G, are shown in the figure \ref{fig:mod_time}. The energy where the synchrotron will dominate over $pp$ interaction, $E'^b_p = 2.95 \times 10^{18}$ eV. We obtained this for the ambient parameters, $n_{H,0} = 3.8 \times 10^{10}$ and $\alpha_d= 0.4$. These values are obtained such that we could explain both neutrino flux as well as VHE events. The typical proton energy required for producing synchrotron photons of critical frequency, $\nu_c$ in the observer frame is,
\begin{equation}
    E'_p=4.38 \times 10^{19} \textup{eV} \, \left(\frac{\nu_{c,25}}{B_3}\right)^{1/2} \, \frac{ (1+z)}{\delta},
\end{equation}
\\ 
where $\nu_{c,25} = \frac{\nu_c}{10^{25}}$. The blue-shaded region shows the energy range of the protons in the jet frame corresponding to the observed VHE $\gamma$-ray event range. The $t_{obs_{psync}}$ for the energy range $E_p' = 1.38 \times 10^{19} eV$ to $4.57 \times 10^{19} eV$ is 44 to 11 days, as shown with the horizontal (red) shaded region. The emissions above the mentioned $E_p'$ get suppressed by the EBL. 

Our model gives a possible explanation for the extended VHE emission from the source, with the varying ambient profile. 

\section{Acknowledgement}
We thank Nayantara Gupta for her valuable comments on using the GAMERA package. Sunanda would like to acknowledge the CSIR for providing a fellowship. RM would like to acknowledge the SERB start-up grant SRG/2020/001932.

\bibliography{apssamp}

\providecommand{\noopsort}[1]{}\providecommand{\singleletter}[1]{#1}%
\begin{thebibliography}{27}%
\makeatletter
\providecommand \@ifxundefined [1]{%
 \@ifx{#1\undefined}
}%
\providecommand \@ifnum [1]{%
 \ifnum #1\expandafter \@firstoftwo
 \else \expandafter \@secondoftwo
 \fi
}%
\providecommand \@ifx [1]{%
 \ifx #1\expandafter \@firstoftwo
 \else \expandafter \@secondoftwo
 \fi
}%
\providecommand \natexlab [1]{#1}%
\providecommand \enquote  [1]{``#1''}%
\providecommand \bibnamefont  [1]{#1}%
\providecommand \bibfnamefont [1]{#1}%
\providecommand \citenamefont [1]{#1}%
\providecommand \href@noop [0]{\@secondoftwo}%
\providecommand \href [0]{\begingroup \@sanitize@url \@href}%
\providecommand \@href[1]{\@@startlink{#1}\@@href}%
\providecommand \@@href[1]{\endgroup#1\@@endlink}%
\providecommand \@sanitize@url [0]{\catcode `\\12\catcode `\$12\catcode
  `\&12\catcode `\#12\catcode `\^12\catcode `\_12\catcode `\%12\relax}%
\providecommand \@@startlink[1]{}%
\providecommand \@@endlink[0]{}%
\providecommand \url  [0]{\begingroup\@sanitize@url \@url }%
\providecommand \@url [1]{\endgroup\@href {#1}{\urlprefix }}%
\providecommand \urlprefix  [0]{URL }%
\providecommand \Eprint [0]{\href }%
\providecommand \doibase [0]{https://doi.org/}%
\providecommand \selectlanguage [0]{\@gobble}%
\providecommand \bibinfo  [0]{\@secondoftwo}%
\providecommand \bibfield  [0]{\@secondoftwo}%
\providecommand \translation [1]{[#1]}%
\providecommand \BibitemOpen [0]{}%
\providecommand \bibitemStop [0]{}%
\providecommand \bibitemNoStop [0]{.\EOS\space}%
\providecommand \EOS [0]{\spacefactor3000\relax}%
\providecommand \BibitemShut  [1]{\csname bibitem#1\endcsname}%
\let\auto@bib@innerbib\@empty
\bibitem [{\citenamefont {Yasuyuki}\ \emph {et~al.}(2017)\citenamefont
  {Yasuyuki}, \citenamefont {{Tanaka}}, \citenamefont {{(Hiroshima
  University)}}, \citenamefont {Sara}, \citenamefont {Buson}, \citenamefont {{
  (NASA/GSFC)}},\ and\ \citenamefont {{Kocevski }(NASA/MSFC)}}]{Fermi_atel}%
  \BibitemOpen
  \bibfield  {author} {\bibinfo {author} {\bibnamefont {Yasuyuki}}, \bibinfo
  {author} {\bibfnamefont {T.}~\bibnamefont {{Tanaka}}}, \bibinfo {author}
  {\bibnamefont {{(Hiroshima University)}}}, \bibinfo {author} {\bibnamefont
  {Sara}}, \bibinfo {author} {\bibnamefont {Buson}}, \bibinfo {author}
  {\bibnamefont {{ (NASA/GSFC)}}},\ and\ \bibinfo {author} {\bibfnamefont
  {D.}~\bibnamefont {{Kocevski }(NASA/MSFC)}},\ }\bibfield  {title} {\bibinfo
  {title} {The astronomer's telegram},\ }\href@noop {} {\ \textbf {\bibinfo
  {volume} {{ 10791}}} (\bibinfo {year} {2017})}\BibitemShut {NoStop}%
\bibitem [{\citenamefont {(SSDC/ASI}\ \emph {et~al.}(2017)\citenamefont
  {(SSDC/ASI}, \citenamefont {INAF/OAR)}, \citenamefont {(INAF/IAPS)},
  \citenamefont {Pittori}, \citenamefont {{(SSDC/ASI and INAF/OAR)}} \emph
  {et~al.}}]{AGILe}%
  \BibitemOpen
  \bibfield  {author} {\bibinfo {author} {\bibfnamefont {F.~L.}\ \bibnamefont
  {(SSDC/ASI}}, \bibinfo {author} {\bibnamefont {INAF/OAR)}}, \bibinfo {author}
  {\bibfnamefont {G.~P.}\ \bibnamefont {(INAF/IAPS)}}, \bibinfo {author}
  {\bibfnamefont {C.}~\bibnamefont {Pittori}}, \bibinfo {author} {\bibfnamefont
  {F.~V.}\ \bibnamefont {{(SSDC/ASI and INAF/OAR)}}}, \emph {et~al.},\
  }\bibfield  {title} {\bibinfo {title} {The astronomer's telegram},\
  }\href@noop {} {\ \textbf {\bibinfo {volume} {10801}} (\bibinfo {year}
  {2017})}\BibitemShut {NoStop}%
\bibitem [{\citenamefont {(DESY)}\ \emph {et~al.}(2017)\citenamefont {(DESY)},
  \citenamefont {Stanek}, \citenamefont {Kochanek}, \citenamefont {(OSU)},
  \citenamefont {Holoien}, \citenamefont {Observatories)},\ and\ \citenamefont
  {P}}]{ASAS}%
  \BibitemOpen
  \bibfield  {author} {\bibinfo {author} {\bibfnamefont {A.~F.}\ \bibnamefont
  {(DESY)}}, \bibinfo {author} {\bibfnamefont {K.~Z.}\ \bibnamefont {Stanek}},
  \bibinfo {author} {\bibfnamefont {C.~S.}\ \bibnamefont {Kochanek}}, \bibinfo
  {author} {\bibfnamefont {T.~A.~T.}\ \bibnamefont {(OSU)}}, \bibinfo {author}
  {\bibfnamefont {T.~W.-S.}\ \bibnamefont {Holoien}}, \bibinfo {author}
  {\bibfnamefont {B.~J. S.~C.}\ \bibnamefont {Observatories)}},\ and\ \bibinfo
  {author} {\bibfnamefont {J.~L.}\ \bibnamefont {P}},\ }\bibfield  {title}
  {\bibinfo {title} {The astronomer's telegram},\ }\href@noop {} {\ \textbf
  {\bibinfo {volume} {10794}} (\bibinfo {year} {2017})}\BibitemShut {NoStop}%
\bibitem [{\citenamefont {de~Naurois}(2017)}]{HESS}%
  \BibitemOpen
  \bibfield  {author} {\bibinfo {author} {\bibfnamefont {M.}~\bibnamefont
  {de~Naurois}},\ }\bibfield  {title} {\bibinfo {title} {The astronomer's
  telegram},\ }\href@noop {} {\ \textbf {\bibinfo {volume} {10787}} (\bibinfo
  {year} {2017})}\BibitemShut {NoStop}%
\bibitem [{\citenamefont {Mukherjee}(2017)}]{veritas}%
  \BibitemOpen
  \bibfield  {author} {\bibinfo {author} {\bibfnamefont {R.}~\bibnamefont
  {Mukherjee}},\ }\bibfield  {title} {\bibinfo {title} {The astronomer's
  telegram},\ }\href@noop {} {\ \textbf {\bibinfo {volume} {10833}} (\bibinfo
  {year} {2017})}\BibitemShut {NoStop}%
\bibitem [{\citenamefont {Martinez}\ \emph {et~al.}(2017)\citenamefont
  {Martinez}, \citenamefont {Taboada}, \citenamefont {Hui},\ and\ \citenamefont
  {Lauer}}]{Hawc}%
  \BibitemOpen
  \bibfield  {author} {\bibinfo {author} {\bibfnamefont {I.}~\bibnamefont
  {Martinez}}, \bibinfo {author} {\bibfnamefont {I.}~\bibnamefont {Taboada}},
  \bibinfo {author} {\bibfnamefont {M.}~\bibnamefont {Hui}},\ and\ \bibinfo
  {author} {\bibfnamefont {R.}~\bibnamefont {Lauer}},\ }\bibfield  {title}
  {\bibinfo {title} {The astronomer's telegram},\ }\href@noop {} {\ \textbf
  {\bibinfo {volume} {10802}} (\bibinfo {year} {2017})}\BibitemShut {NoStop}%
\bibitem [{\citenamefont {Mirzoyan}(2017)}]{Magic}%
  \BibitemOpen
  \bibfield  {author} {\bibinfo {author} {\bibfnamefont {R.}~\bibnamefont
  {Mirzoyan}},\ }\bibfield  {title} {\bibinfo {title} {The astronomer's
  telegram},\ }\href@noop {} {\ \textbf {\bibinfo {volume} {10817}} (\bibinfo
  {year} {2017})}\BibitemShut {NoStop}%
\bibitem [{\citenamefont {Ansoldi}\ \emph {et~al.}(2018)\citenamefont
  {Ansoldi}, \citenamefont {Antonelli}, \citenamefont {Arcaro}, \citenamefont
  {Baack}, \citenamefont {Babi{\'c}}, \citenamefont {Banerjee}, \citenamefont
  {Bangale}, \citenamefont {Barres~de Almeida}, \citenamefont {Barrio} \emph
  {et~al.}}]{2018ApJ...863L..10A}%
  \BibitemOpen
  \bibfield  {author} {\bibinfo {author} {\bibfnamefont {S.}~\bibnamefont
  {Ansoldi}}, \bibinfo {author} {\bibfnamefont {L.~A.}\ \bibnamefont
  {Antonelli}}, \bibinfo {author} {\bibfnamefont {C.}~\bibnamefont {Arcaro}},
  \bibinfo {author} {\bibfnamefont {D.}~\bibnamefont {Baack}}, \bibinfo
  {author} {\bibfnamefont {A.}~\bibnamefont {Babi{\'c}}}, \bibinfo {author}
  {\bibfnamefont {B.}~\bibnamefont {Banerjee}}, \bibinfo {author}
  {\bibfnamefont {P.}~\bibnamefont {Bangale}}, \bibinfo {author} {\bibfnamefont
  {U.}~\bibnamefont {Barres~de Almeida}}, \bibinfo {author} {\bibfnamefont
  {J.~A.}\ \bibnamefont {Barrio}}, \emph {et~al.},\ }\bibfield  {title}
  {\bibinfo {title} {{The Blazar TXS 0506+056 Associated with a High-energy
  Neutrino: Insights into Extragalactic Jets and Cosmic-Ray Acceleration}},\
  }\href {https://doi.org/10.3847/2041-8213/aad083} {\bibfield  {journal}
  {\bibinfo  {journal} {The Astrophysical Journal Letter}\ }\textbf {\bibinfo
  {volume} {863}},\ \bibinfo {eid} {L10} (\bibinfo {year} {2018})},\ \Eprint
  {https://arxiv.org/abs/1807.04300} {arXiv:1807.04300 [astro-ph.HE]}
  \BibitemShut {NoStop}%
\bibitem [{\citenamefont {Acciari}\ \emph {et~al.}(2022)\citenamefont
  {Acciari}, \citenamefont {Aniello}, \citenamefont {{Ansoldi}}, \citenamefont
  {Antonelli} \emph {et~al.}}]{2022ApJ...927..197A}%
  \BibitemOpen
  \bibfield  {author} {\bibinfo {author} {\bibfnamefont {V.~A.}\ \bibnamefont
  {Acciari}}, \bibinfo {author} {\bibfnamefont {T.}~\bibnamefont {Aniello}},
  \bibinfo {author} {\bibfnamefont {S.}~\bibnamefont {{Ansoldi}}}, \bibinfo
  {author} {\bibfnamefont {L.~A.}\ \bibnamefont {Antonelli}}, \emph {et~al.},\
  }\bibfield  {title} {\bibinfo {title} {{Investigating the Blazar TXS 0506+056
  through Sharp Multiwavelength Eyes During 2017-2019}},\ }\href
  {https://doi.org/10.3847/1538-4357/ac531d} {\bibfield  {journal} {\bibinfo
  {journal} {\apj}\ }\textbf {\bibinfo {volume} {927}},\ \bibinfo {eid} {197}
  (\bibinfo {year} {2022})},\ \Eprint {https://arxiv.org/abs/2202.02600}
  {arXiv:2202.02600 [astro-ph.HE]} \BibitemShut {NoStop}%
\bibitem [{\citenamefont {Gao}\ \emph {et~al.}(2019)\citenamefont {Gao},
  \citenamefont {Fedynitch}, \citenamefont {Winter},\ and\ \citenamefont
  {Pohl}}]{Gao:2018mnu}%
  \BibitemOpen
  \bibfield  {author} {\bibinfo {author} {\bibfnamefont {S.}~\bibnamefont
  {Gao}}, \bibinfo {author} {\bibfnamefont {A.}~\bibnamefont {Fedynitch}},
  \bibinfo {author} {\bibfnamefont {W.}~\bibnamefont {Winter}},\ and\ \bibinfo
  {author} {\bibfnamefont {M.}~\bibnamefont {Pohl}},\ }\bibfield  {title}
  {\bibinfo {title} {{Modelling the coincident observation of a high-energy
  neutrino and a bright blazar flare}},\ }\href
  {https://doi.org/10.1038/s41550-018-0610-1} {\bibfield  {journal} {\bibinfo
  {journal} {Nature Astron.}\ }\textbf {\bibinfo {volume} {3}},\ \bibinfo
  {pages} {88} (\bibinfo {year} {2019})},\ \Eprint
  {https://arxiv.org/abs/1807.04275} {arXiv:1807.04275 [astro-ph.HE]}
  \BibitemShut {NoStop}%
\bibitem [{\citenamefont {Keivani}\ \emph {et~al.}(2018)\citenamefont {Keivani}
  \emph {et~al.}}]{Keivani:2018rnh}%
  \BibitemOpen
  \bibfield  {author} {\bibinfo {author} {\bibfnamefont {A.}~\bibnamefont
  {Keivani}} \emph {et~al.},\ }\bibfield  {title} {\bibinfo {title} {{A
  Multimessenger Picture of the Flaring Blazar TXS 0506+056: implications for
  High-Energy Neutrino Emission and Cosmic Ray Acceleration}},\ }\href
  {https://doi.org/10.3847/1538-4357/aad59a} {\bibfield  {journal} {\bibinfo
  {journal} {Astrophys. J.}\ }\textbf {\bibinfo {volume} {864}},\ \bibinfo
  {pages} {84} (\bibinfo {year} {2018})},\ \Eprint
  {https://arxiv.org/abs/1807.04537} {arXiv:1807.04537 [astro-ph.HE]}
  \BibitemShut {NoStop}%
\bibitem [{\citenamefont {Cerruti}\ \emph {et~al.}(2019)\citenamefont
  {Cerruti}, \citenamefont {Zech}, \citenamefont {Boisson}, \citenamefont
  {Emery}, \citenamefont {Inoue},\ and\ \citenamefont
  {Lenain}}]{Cerruti:2018tmc}%
  \BibitemOpen
  \bibfield  {author} {\bibinfo {author} {\bibfnamefont {M.}~\bibnamefont
  {Cerruti}}, \bibinfo {author} {\bibfnamefont {A.}~\bibnamefont {Zech}},
  \bibinfo {author} {\bibfnamefont {C.}~\bibnamefont {Boisson}}, \bibinfo
  {author} {\bibfnamefont {G.}~\bibnamefont {Emery}}, \bibinfo {author}
  {\bibfnamefont {S.}~\bibnamefont {Inoue}},\ and\ \bibinfo {author}
  {\bibfnamefont {J.~P.}\ \bibnamefont {Lenain}},\ }\bibfield  {title}
  {\bibinfo {title} {{Leptohadronic single-zone models for the electromagnetic
  and neutrino emission of TXS 0506+056}},\ }\href
  {https://doi.org/10.1093/mnrasl/sly210} {\bibfield  {journal} {\bibinfo
  {journal} {Mon. Not. Roy. Astron. Soc.}\ }\textbf {\bibinfo {volume} {483}},\
  \bibinfo {pages} {L12} (\bibinfo {year} {2019})},\ \bibinfo {note} {[Erratum:
  Mon.Not.Roy.Astron.Soc. 502, L21--L22 (2021)]},\ \Eprint
  {https://arxiv.org/abs/1807.04335} {arXiv:1807.04335 [astro-ph.HE]}
  \BibitemShut {NoStop}%
\bibitem [{\citenamefont {Oikonomou}\ \emph {et~al.}(2019)\citenamefont
  {Oikonomou}, \citenamefont {Murase}, \citenamefont {Padovani}, \citenamefont
  {Resconi},\ and\ \citenamefont {M\'esz\'aros}}]{Oikonomou:2019djc}%
  \BibitemOpen
  \bibfield  {author} {\bibinfo {author} {\bibfnamefont {F.}~\bibnamefont
  {Oikonomou}}, \bibinfo {author} {\bibfnamefont {K.}~\bibnamefont {Murase}},
  \bibinfo {author} {\bibfnamefont {P.}~\bibnamefont {Padovani}}, \bibinfo
  {author} {\bibfnamefont {E.}~\bibnamefont {Resconi}},\ and\ \bibinfo {author}
  {\bibfnamefont {P.}~\bibnamefont {M\'esz\'aros}},\ }\bibfield  {title}
  {\bibinfo {title} {{High energy neutrino flux from individual blazar
  flares}},\ }\href {https://doi.org/10.1093/mnras/stz2246} {\bibfield
  {journal} {\bibinfo  {journal} {Mon. Not. Roy. Astron. Soc.}\ }\textbf
  {\bibinfo {volume} {489}},\ \bibinfo {pages} {4347} (\bibinfo {year}
  {2019})},\ \Eprint {https://arxiv.org/abs/1906.05302} {arXiv:1906.05302
  [astro-ph.HE]} \BibitemShut {NoStop}%
\bibitem [{\citenamefont {Petropoulou}\ \emph {et~al.}(2020)\citenamefont
  {Petropoulou} \emph {et~al.}}]{Petropoulou:2019zqp}%
  \BibitemOpen
  \bibfield  {author} {\bibinfo {author} {\bibfnamefont {M.}~\bibnamefont
  {Petropoulou}} \emph {et~al.},\ }\bibfield  {title} {\bibinfo {title}
  {{Multi-Epoch Modeling of TXS 0506+056 and Implications for Long-Term
  High-Energy Neutrino Emission}},\ }\href
  {https://doi.org/10.3847/1538-4357/ab76d0} {\bibfield  {journal} {\bibinfo
  {journal} {Astrophys. J.}\ }\textbf {\bibinfo {volume} {891}},\ \bibinfo
  {pages} {115} (\bibinfo {year} {2020})},\ \Eprint
  {https://arxiv.org/abs/1911.04010} {arXiv:1911.04010 [astro-ph.HE]}
  \BibitemShut {NoStop}%
\bibitem [{\citenamefont {Murase}\ \emph {et~al.}(2018)\citenamefont {Murase},
  \citenamefont {Oikonomou},\ and\ \citenamefont
  {Petropoulou}}]{Murase:2018iyl}%
  \BibitemOpen
  \bibfield  {author} {\bibinfo {author} {\bibfnamefont {K.}~\bibnamefont
  {Murase}}, \bibinfo {author} {\bibfnamefont {F.}~\bibnamefont {Oikonomou}},\
  and\ \bibinfo {author} {\bibfnamefont {M.}~\bibnamefont {Petropoulou}},\
  }\bibfield  {title} {\bibinfo {title} {{Blazar Flares as an Origin of
  High-Energy Cosmic Neutrinos?}},\ }\href
  {https://doi.org/10.3847/1538-4357/aada00} {\bibfield  {journal} {\bibinfo
  {journal} {Astrophys. J.}\ }\textbf {\bibinfo {volume} {865}},\ \bibinfo
  {pages} {124} (\bibinfo {year} {2018})},\ \Eprint
  {https://arxiv.org/abs/1807.04748} {arXiv:1807.04748 [astro-ph.HE]}
  \BibitemShut {NoStop}%
\bibitem [{\citenamefont {Sahakyan}(2018)}]{Sahakyan:2018voh}%
  \BibitemOpen
  \bibfield  {author} {\bibinfo {author} {\bibfnamefont {N.}~\bibnamefont
  {Sahakyan}},\ }\bibfield  {title} {\bibinfo {title} {{Lepto-hadronic
  $\gamma$-ray and neutrino emission from the jet of TXS 0506+056}},\ }\href
  {https://doi.org/10.3847/1538-4357/aadade} {\bibfield  {journal} {\bibinfo
  {journal} {Astrophys. J.}\ }\textbf {\bibinfo {volume} {866}},\ \bibinfo
  {pages} {109} (\bibinfo {year} {2018})},\ \Eprint
  {https://arxiv.org/abs/1808.05651} {arXiv:1808.05651 [astro-ph.HE]}
  \BibitemShut {NoStop}%
\bibitem [{\citenamefont {Liu}\ \emph {et~al.}(2019)\citenamefont {Liu},
  \citenamefont {Wang}, \citenamefont {Xue}, \citenamefont {Taylor},
  \citenamefont {Wang}, \citenamefont {Li},\ and\ \citenamefont
  {Yan}}]{Liu:2018utd}%
  \BibitemOpen
  \bibfield  {author} {\bibinfo {author} {\bibfnamefont {R.-Y.}\ \bibnamefont
  {Liu}}, \bibinfo {author} {\bibfnamefont {K.}~\bibnamefont {Wang}}, \bibinfo
  {author} {\bibfnamefont {R.}~\bibnamefont {Xue}}, \bibinfo {author}
  {\bibfnamefont {A.~M.}\ \bibnamefont {Taylor}}, \bibinfo {author}
  {\bibfnamefont {X.-Y.}\ \bibnamefont {Wang}}, \bibinfo {author}
  {\bibfnamefont {Z.}~\bibnamefont {Li}},\ and\ \bibinfo {author}
  {\bibfnamefont {H.}~\bibnamefont {Yan}},\ }\bibfield  {title} {\bibinfo
  {title} {{Hadronuclear interpretation of a high-energy neutrino event
  coincident with a blazar flare}},\ }\href
  {https://doi.org/10.1103/PhysRevD.99.063008} {\bibfield  {journal} {\bibinfo
  {journal} {Phys. Rev. D}\ }\textbf {\bibinfo {volume} {99}},\ \bibinfo
  {pages} {063008} (\bibinfo {year} {2019})},\ \Eprint
  {https://arxiv.org/abs/1807.05113} {arXiv:1807.05113 [astro-ph.HE]}
  \BibitemShut {NoStop}%
\bibitem [{\citenamefont {Banik}\ and\ \citenamefont
  {Bhadra}(2019)}]{Banik:2019jlm}%
  \BibitemOpen
  \bibfield  {author} {\bibinfo {author} {\bibfnamefont {P.}~\bibnamefont
  {Banik}}\ and\ \bibinfo {author} {\bibfnamefont {A.}~\bibnamefont {Bhadra}},\
  }\bibfield  {title} {\bibinfo {title} {{Describing correlated observations of
  neutrinos and gamma-ray flares from the blazar TXS 0506+056 with a proton
  blazar model}},\ }\href {https://doi.org/10.1103/PhysRevD.99.103006}
  {\bibfield  {journal} {\bibinfo  {journal} {Phys. Rev. D}\ }\textbf {\bibinfo
  {volume} {99}},\ \bibinfo {pages} {103006} (\bibinfo {year} {2019})},\
  \Eprint {https://arxiv.org/abs/1908.11849} {arXiv:1908.11849 [astro-ph.HE]}
  \BibitemShut {NoStop}%
\bibitem [{\citenamefont {Banik}\ \emph {et~al.}(2020)\citenamefont {Banik},
  \citenamefont {Bhadra}, \citenamefont {Pandey},\ and\ \citenamefont
  {Majumdar}}]{Banik:2019twt}%
  \BibitemOpen
  \bibfield  {author} {\bibinfo {author} {\bibfnamefont {P.}~\bibnamefont
  {Banik}}, \bibinfo {author} {\bibfnamefont {A.}~\bibnamefont {Bhadra}},
  \bibinfo {author} {\bibfnamefont {M.}~\bibnamefont {Pandey}},\ and\ \bibinfo
  {author} {\bibfnamefont {D.}~\bibnamefont {Majumdar}},\ }\bibfield  {title}
  {\bibinfo {title} {{Implications of a proton blazar inspired model on
  correlated observations of neutrinos with gamma-ray flaring blazars}},\
  }\href {https://doi.org/10.1103/PhysRevD.101.063024} {\bibfield  {journal}
  {\bibinfo  {journal} {Phys. Rev. D}\ }\textbf {\bibinfo {volume} {101}},\
  \bibinfo {pages} {063024} (\bibinfo {year} {2020})},\ \Eprint
  {https://arxiv.org/abs/1909.01993} {arXiv:1909.01993 [astro-ph.HE]}
  \BibitemShut {NoStop}%
\bibitem [{\citenamefont {Tetarenko}\ \emph {et~al.}(2017)\citenamefont
  {Tetarenko}, \citenamefont {(UAlberta)}, \citenamefont {(NRAO)},\ and\
  \citenamefont {(Curtin-ICRAR)}}]{VLA-2017}%
  \BibitemOpen
  \bibfield  {author} {\bibinfo {author} {\bibfnamefont {A.~J.}\ \bibnamefont
  {Tetarenko}}, \bibinfo {author} {\bibfnamefont {G.~R.~S.}\ \bibnamefont
  {(UAlberta)}}, \bibinfo {author} {\bibfnamefont {A.~E.~K.}\ \bibnamefont
  {(NRAO)}},\ and\ \bibinfo {author} {\bibfnamefont {J.~C. M.-J.}\ \bibnamefont
  {(Curtin-ICRAR)}},\ }\bibfield  {title} {\bibinfo {title} {The astronomer's
  telegram},\ }\href@noop {} {\ \textbf {\bibinfo {volume} {10861}} (\bibinfo
  {year} {2017})}\BibitemShut {NoStop}%
\bibitem [{\citenamefont {Yamanaka}\ \emph {et~al.}(2017)\citenamefont
  {Yamanaka}, \citenamefont {Tanaka}, \citenamefont {Mori} \emph
  {et~al.}}]{kanata-2017}%
  \BibitemOpen
  \bibfield  {author} {\bibinfo {author} {\bibfnamefont {M.}~\bibnamefont
  {Yamanaka}}, \bibinfo {author} {\bibfnamefont {Y.~T.}\ \bibnamefont
  {Tanaka}}, \bibinfo {author} {\bibfnamefont {H.}~\bibnamefont {Mori}}, \emph
  {et~al.},\ }\bibfield  {title} {\bibinfo {title} {The astronomer's
  telegram},\ }\href@noop {} {\ \textbf {\bibinfo {volume} {10844}} (\bibinfo
  {year} {2017})}\BibitemShut {NoStop}%
\bibitem [{\citenamefont {Keivani}\ \emph {et~al.}(2017)\citenamefont
  {Keivani}, \citenamefont {(PSU)}, \citenamefont {{P.A.}}, \citenamefont
  {{Evans}}, \citenamefont {{Leicester}} \emph {et~al.}}]{Swift-XRT}%
  \BibitemOpen
  \bibfield  {author} {\bibinfo {author} {\bibfnamefont {A.}~\bibnamefont
  {Keivani}}, \bibinfo {author} {\bibnamefont {(PSU)}}, \bibinfo {author}
  {\bibnamefont {{P.A.}}}, \bibinfo {author} {\bibnamefont {{Evans}}}, \bibinfo
  {author} {\bibfnamefont {U.}~\bibnamefont {{Leicester}}}, \emph {et~al.},\
  }\bibfield  {title} {\bibinfo {title} {Gcn circular},\ }\href@noop {} {\
  \textbf {\bibinfo {volume} {{ 21930}}} (\bibinfo {year} {2017})}\BibitemShut
  {NoStop}%
\bibitem [{\citenamefont {(PSU)}\ \emph {et~al.}(2017)\citenamefont {(PSU)},
  \citenamefont {(PSU)}, \citenamefont {(PSU)} \emph {et~al.}}]{Nustar}%
  \BibitemOpen
  \bibfield  {author} {\bibinfo {author} {\bibfnamefont {D.~B.~F.}\
  \bibnamefont {(PSU)}}, \bibinfo {author} {\bibfnamefont {J.~J.~D.}\
  \bibnamefont {(PSU)}}, \bibinfo {author} {\bibfnamefont {A.~K.}\ \bibnamefont
  {(PSU)}}, \emph {et~al.},\ }\bibfield  {title} {\bibinfo {title} {The
  astronomer's telegram},\ }\href@noop {} {\ \textbf {\bibinfo {volume}
  {10845}} (\bibinfo {year} {2017})}\BibitemShut {NoStop}%
\bibitem [{\citenamefont {Abdo}\ \emph {et~al.}(2010)\citenamefont {Abdo},
  \citenamefont {Ackermann}, \citenamefont {Ajello}, \citenamefont {Antolini}
  \emph {et~al.}}]{2010ApJ...722..520A}%
  \BibitemOpen
  \bibfield  {author} {\bibinfo {author} {\bibfnamefont {A.~A.}\ \bibnamefont
  {Abdo}}, \bibinfo {author} {\bibfnamefont {M.}~\bibnamefont {Ackermann}},
  \bibinfo {author} {\bibfnamefont {M.}~\bibnamefont {Ajello}}, \bibinfo
  {author} {\bibfnamefont {E.}~\bibnamefont {Antolini}}, \emph {et~al.},\
  }\bibfield  {title} {\bibinfo {title} {{Gamma-ray Light Curves and
  Variability of Bright Fermi-detected Blazars}},\ }\href
  {https://doi.org/10.1088/0004-637X/722/1/520} {\bibfield  {journal} {\bibinfo
   {journal} {\apj}\ }\textbf {\bibinfo {volume} {722}},\ \bibinfo {pages}
  {520} (\bibinfo {year} {2010})},\ \Eprint {https://arxiv.org/abs/1004.0348}
  {arXiv:1004.0348 [astro-ph.HE]} \BibitemShut {NoStop}%
\bibitem [{\citenamefont {Hahn}(2016)}]{Hahn:2015hhw}%
  \BibitemOpen
  \bibfield  {author} {\bibinfo {author} {\bibfnamefont {J.}~\bibnamefont
  {Hahn}},\ }\bibfield  {title} {\bibinfo {title} {{GAMERA - a new modeling
  package for non-thermal spectral modeling}},\ }\href
  {https://doi.org/10.22323/1.236.0917} {\bibfield  {journal} {\bibinfo
  {journal} {PoS}\ }\textbf {\bibinfo {volume} {ICRC2015}},\ \bibinfo {pages}
  {917} (\bibinfo {year} {2016})}\BibitemShut {NoStop}%
\bibitem [{\citenamefont {{Kelner}}\ \emph {et~al.}(2006)\citenamefont
  {{Kelner}}, \citenamefont {{Aharonian}},\ and\ \citenamefont
  {{Bugayov}}}]{2006PhRvD..74c4018K}%
  \BibitemOpen
  \bibfield  {author} {\bibinfo {author} {\bibfnamefont {S.~R.}\ \bibnamefont
  {{Kelner}}}, \bibinfo {author} {\bibfnamefont {F.~A.}\ \bibnamefont
  {{Aharonian}}},\ and\ \bibinfo {author} {\bibfnamefont {V.~V.}\ \bibnamefont
  {{Bugayov}}},\ }\bibfield  {title} {\bibinfo {title} {{Energy spectra of
  gamma rays, electrons, and neutrinos produced at proton-proton interactions
  in the very high energy regime}},\ }\href
  {https://doi.org/10.1103/PhysRevD.74.034018} {\bibfield  {journal} {\bibinfo
  {journal} {\prd}\ }\textbf {\bibinfo {volume} {74}},\ \bibinfo {eid} {034018}
  (\bibinfo {year} {2006})},\ \Eprint {https://arxiv.org/abs/astro-ph/0606058}
  {arXiv:astro-ph/0606058 [astro-ph]} \BibitemShut {NoStop}%
\bibitem [{\citenamefont {Ptitsyna}\ and\ \citenamefont
  {Troitsky}(2010)}]{Ptitsyna:2008zs}%
  \BibitemOpen
  \bibfield  {author} {\bibinfo {author} {\bibfnamefont {K.~V.}\ \bibnamefont
  {Ptitsyna}}\ and\ \bibinfo {author} {\bibfnamefont {S.~V.}\ \bibnamefont
  {Troitsky}},\ }\bibfield  {title} {\bibinfo {title} {{Physical conditions in
  potential sources of ultra-high-energy cosmic rays. I. Updated Hillas plot
  and radiation-loss constraints}},\ }\href
  {https://doi.org/10.3367/UFNe.0180.201007c.0723} {\bibfield  {journal}
  {\bibinfo  {journal} {Phys. Usp.}\ }\textbf {\bibinfo {volume} {53}},\
  \bibinfo {pages} {691} (\bibinfo {year} {2010})},\ \Eprint
  {https://arxiv.org/abs/0808.0367} {arXiv:0808.0367 [astro-ph]} \BibitemShut
  {NoStop}%
\end{thebibliography}%

\end{document}